# Strong spin-orbit interaction and magnetotransport in semiconductor Bi$_2$O$_2$Se nanoplates

Mengmeng Meng,[1] Shaoyun Huang,[1,*] Congwei Tan,[2] Jinxiong Wu,[2] Yumei Jing,[1] Hailin Peng[2,*] and H. Q. Xu[1,3,*]

[1]Beijing Key Laboratory of Quantum Devices, Key Laboratory for the Physics and Chemistry of Nanodevices, and Department of Electronics, Peking University, Beijing 100871, China

[2]Center for Nanochemistry, Beijing National Laboratory for Molecular Sciences (BNLMS), College of Chemistry and Molecular Engineering, Peking University, Beijing 100871, China.

[3]Division of Solid State Physics, Lund University, Box 118, S-221 00 Lund, Sweden

Semiconductor Bi$_2$O$_2$Se nanolayers of high crystal quality have been realized via epitaxial growth. These two-dimensional (2D) materials possess excellent electron transport properties with potential application in nanoelectronics. It is also strongly expected that the 2D Bi$_2$O$_2$Se nanolayers could be of an excellent material platform for developing spintronic and topological quantum devices, if the presence of strong spin-orbit interaction in the 2D materials can be experimentally demonstrated. Here, we report on experimental determination of the strength of spin-orbit interaction in Bi$_2$O$_2$Se nanoplates through magnetotransport measurements. The nanoplates are epitaxially grown by chemical vapor deposition and the magnetotransport measurements are performed at low temperatures. The measured magnetoconductance exhibits a crossover behavior from weak antilocalization to weak localization at low magnetic fields with increasing temperature or decreasing back gate voltage. We have analyzed this transition behavior of the magnetoconductance based on an interference theory which describes the quantum correction to the magnetoconductance of a 2D system in the presence of spin-orbit interaction. Dephasing length and spin relaxation length are extracted from the magnetoconductance measurements. Comparing to other semiconductor nanostructures, the extracted relatively short spin relaxation length of ~150 nm indicates the existence of strong spin-orbit interaction in Bi$_2$O$_2$Se nanolayers.

The discovery of graphene has inspired extensive investigation of electrical, mechanical and optical properties of two-dimensional (2D) materials[1-5]. As one of emerging 2D semiconducting materials, high-quality $Bi_2O_2Se$ nanoplates has recently been successfully grown by chemical vapor deposition (CVD) on mica substrates[6]. The grown 2D $Bi_2O_2Se$ nanoplates not only possess a band energy gap, which is desired for applications in planar nanoelectronic logic devices and circuits, but also are stable against oxidation and moisture in the air[6,7]. The electron field-effect mobility of the nanoplates is found to reach ~450 $cm^2$/V·s at room temperature. At the low temperature of 2 K, the high electron Hall mobility of ~20000 $cm^2$/V·s is detected and Shubnikov-de Haas (SdH) quantum coherent oscillations are observed at moderate magnetic fields in the $Bi_2O_2Se$ nanoplates[6], implying that the materials have excellent transport properties and could be potentially used as a new host material system for exploration of novel quantum transport phenomena. Considering the conduction electron states and the valence hole states around the band gap are dominantly built from the p-orbitals of Bi and Se elements[6], 2D $Bi_2O_2Se$ nanoplates are expected to possess strong spin-orbit interaction. Such layered semiconducting, strong-spin interacting electron systems are desired for constructing planar topological superconducting systems, in which Majorana bound states[8-13] can be created and manipulated, with potential applications in topological quantum computing. Moreover, since electron spins can be manipulated by an electric field in an electron system with strong spin-orbit interaction, 2D $Bi_2O_2Se$ nanoplates could be utilized for developing spintronic field-effect transistors and spin-based quantum information technology[14-16]. Thus, the main purpose of this work is to experimentally determine the strength of spin-orbit interaction in 2D $Bi_2O_2Se$ nanoplates through magnetotransport measurements.

Magnetotransport is a key method to determine characteristic length scales in mesoscopic systems[17], including spin relaxation length ($L_{so}$), which is important to understand and control transport and spin manipulation process. In a scattered, time reversal electron system where electrons move in a diffusive instead of ballistic way, quantum interference occurs on closed paths, making a correction to the classical conductivity. With applying a small magnetic field, due to the suppression of the constructive interference between time-reversed closed trajectories, magnetoconductance increases with increasing magnetic field and displays the weak localization (WL) characteristics. In the presence of strong spin-orbit interaction, quantum correction to the zero-magnetic field classical conductivity becomes positive, leading to decrease in magnetoconductance with increasing magnetic field and thus the weak antilocalization (WAL) characteristics[18].

Here, we report on a detailed magnetotransport study of a $Bi_2O_2Se$ nanoplate field-effect device. The low-field magnetoconductance has been measured as a function of temperature and voltage applied to the back gate, and the results are analyzed by means of the Hikami–Larkin–Nagaoka (HLN) interference model. The electron phase coherence length and spin relaxation length in the measured $Bi_2O_2Se$ nanoplate are extracted. Our experiment demonstrates the presence of a strong spin-orbit interaction in the $Bi_2O_2Se$ nanoplate. A crossover between the WAL to WL characteristics with tuning temperature or back gate voltage has also been observed in the $Bi_2O_2Se$ nanoplate. The analysis shows that the crossover is governed by the relative scales of the electron phase coherence length and spin relaxation length. This work demonstrates for the first time an experimental observation of strong spin-orbit interaction in a 2D $Bi_2O_2Se$ nanoplate and will stimulate experimental and theoretical studies of the materials and their potential applications.

The $Bi_2O_2Se$ nanoplates used in this work are grown in a homemade low-pressure CVD system with $Bi_2O_3$ powder and $Bi_2Se_3$ bulk as sources. The $Bi_2O_3$ powder (Alfa Aesar, 99.995%) is placed at the center of a horizontally arranged quartz tube and the $Bi_2Se_3$ bulk (Alfa Aesar, 99.995%) at ~6 cm upstream[19]. Freshly cleaved fluorophlogopite mica is employed as growth substrate, which possesses atomically flat surface and is an ideal substrate to synthesize ultra-thin 2D materials[20, 21]. Argon is used as the carrier gas to transfer the precursor to the growth region. For the further details regarding the materials growth, we refer to Refs. 6 and 20.

Figure 1 shows the structural characterization of the as-grown $Bi_2O_2Se$ nanoplates. $Bi_2O_2Se$ is a typical bismuth-based oxychalcogenide, possessing a tetragonal structure in the I4/mmm space group (a = 3.88 Å, c = 12.16 Å and Z= 2)[22]. $Bi_2O_2Se$ is also an ionic layered material, where positively charged $[Bi_2O_2]_n^{2n+}$ layers are sandwiched by negatively charged $[Se]_n^{2n-}$ layers through relatively weak electrostatic interactions (Figure 1a). Cooperated with appropriate thermodynamics, the strong interlayer/intra-layer bonding anisotropy of $Bi_2O_2Se$ crystal promises to form a 2D crystal on a mica substrate. Figure 1b shows an optical image of some as-grown $Bi_2O_2Se$ nanoplates on a growth mica substrate. As shown in the figure, the nanoplates exhibit a large domain size[6] (about 50×50 μm$^2$ or larger). Square shape is arising from the tetragonal structure of $Bi_2O_2Se$. Figure 1c displays a representative atomic force microscope (AFM) image of a domain of a single $Bi_2O_2Se$ nanoplate, showing that as-grown nanoplates have an atomically flat surface. It has been observed that the atomically flat surfaces of $Bi_2O_2Se$ nanoplates remains almost the same after exposed to the air for months, demonstrating the stability of the oxide semiconductor layers in the ambient environment. The thickness of the nanoplates ranges from several to tens of nanometer, and Figure 1c shows a nanoplate with a thickness of ~6.2 nm (10 layers). The crystalline structure of the as-grown $Bi_2O_2Se$ nanoplates is examined by high-resolution transmission electron microscope (HRTEM) measurements. Figure 1d displays a HRTEM image of an as-grown $Bi_2O_2Se$ nanoplate, where a well-defined arrangement of Bi, O, and Se atomic lattices can be identified, see the area marked by a colored square in the figure. Our HRTEM measurements shows that the CVD grown 2D $Bi_2O_2Se$ nanoplates are of high-quality single crystals.

As-grown $Bi_2O_2Se$ nanoplates are transferred onto a 300-nm-thick $SiO_2$/Si substrate (which is to be used as the back gate) with predefined positioning markers for device fabrication through a PMMA (polymethyl methacrylate)-mediated method.[23] After removal of the mediated PMMA used during the transfer process, the $Bi_2O_2Se$ nanoplates on the device substrate are again covered with PMMA by spin coating. Contact electrodes are then fabricated on selected $Bi_2O_2Se$ nanoplates by electron-beam lithography, electron-beam evaporation of a Ti/Au (5/90 nm) metal layer, and lift-off process. Here, we note that before the metal evaporation, soft Ar plasma is used to remove the residual PMMA resist in the $Bi_2O_2Se$ contact areas to ensure obtaining clean metal-$Bi_2O_2Se$ interfaces. Figure 2a shows an optical image of a device fabricated from a $Bi_2O_2Se$ nanoplate of ~8 nm in thickness. Four parallel metal electrodes with a width of 800 nm and a pitch of 2.6 μm are made on the $Bi_2O_2Se$ nanoplate. Figure 2b shows the schematic structure of the fabricated device and the circuit setup for the magnetotransport measurements presented in this work. The magnetotransport measurements are carried out in a physical property measurement system (PPMS) cryostat equipped with a uniaxial magnet. As shown in Figure 2b, the magnetotransport measurements are performed in a four-probe configuration using a standard lock-in technique, in which a 17-Hz AC current is supplied between the two outer electrodes and the voltage between the two inner contact electrodes is detected. To suppress the influence of Joule heating but in the same time also conductance fluctuations, low-field magnetoresistance measurements at each temperature are performed at several excitation currents. After evaluation of the test measurements, an optimal current excitation of 100 nA is found and is set in the following magnetotransport measurements[20, 24, 25].

Figure 2c shows the measured transfer characteristics of the fabricated $Bi_2O_2Se$

nanoplate device shown in Figure 2a, where the conductance, $G=I/V$, measured using the circuit setup depicted in the inset of Figure 2c at zero magnetic field is plotted as a function of voltage $V_{bg}$ applied to the back gate at different temperatures. It is seen that the device is a typical n-type transistor, and the conduction channel is in an open state at zero back gate voltage and can be switched off by applying a negative gate voltage. At 2 K, the threshold voltage at which the device is switched off is about $-7.2$ V, see Supplementary Information. The threshold voltage moves toward to more negative back gate voltage with increasing temperature. This is a common feature of semiconductor field-effect devices, due to the presence of a high carrier density at a high temperature. At a large positive back gate voltage where the device is at an on-state, the conductance decreases with increasing temperature. This is due to increase in phonon scattering with increasing temperature, as seen in other field–effect devices made from semiconductor nanoplates[26]. Taking into account the device geometry, the field-effect mobility and carrier density can be extracted from the measured transfer characteristics[27], see Supplementary Information. At 2 K, the extracted field-effect mobility and carrier density are 1509 $cm^2/V \cdot s$ and $2.7 \times 10^{12}$ $cm^{-2}$ at $V_{bg} = 30$ V, giving a carrier mean-free path of $l_e \sim 35$ nm in the $Bi_2O_2Se$ nanoplate. This mean-free path value is much smaller than the distance between the two inner probes and thus the carrier transport in our $Bi_2O_2Se$ nanoplate device is in the diffusive regime even at the low temperature of 2 K (see Supplementary Information for the measurements of the carrier density, mobility, and mean free path in the $Bi_2O_2Se$ nanoplate at elevated temperatures).

To determine the electron phase coherence length and spin relaxation length in the $Bi_2O_2Se$ nanoplate, low-field magnetotransport measurements are performed in the fabricated $Bi_2O_2Se$ nanoplate device at low temperatures. Figure 3a shows the measured magnetoconductance $\Delta G = G(B) - G(B = 0)$ at 2 K and at different back gate voltages. Here, the measured curves are successively offset vertically for clarity. It is seen that the measured magnetoconductance shows a WAL-WL crossover with decreasing back gate voltage. At positive back gate voltages, the magnetoconductance near zero magnetic field shows the WAL characteristics, suggesting the existence of strong spin-orbit interaction in the $Bi_2O_2Se$ nanoplate. As the back gate voltage decreases towards negative values, the WAL characteristics are gradually suppressed and the magnetoconductance shows dominantly the WL characteristics. Such a gate-voltage tunable crossover between WAL and WL has also been observed in systems with strong spin-orbit interaction, such as $Al_xGa_{1-x}N/GaN$ heterostructures[28], InAs nanowires[29, 30] and InSb nanowires[31].

In a 2D disorder system, the low-field magnetoconductance is well described by the HLN quantum interference model.[32, 33] Assuming that the transport in the $Bi_2O_2Se$ nanoplate is in the 2D disorder regime, as we will show below this assumption is appropriate, the quantum correction to the low-field magnetoconductance of the device is given by[32, 33]

$$\Delta G(B) = -\frac{e^2}{\pi h}\left[\frac{1}{2}\Psi\left(\frac{B_\phi}{B}+\frac{1}{2}\right) + \Psi\left(\frac{B_{so}+B_e}{B}+\frac{1}{2}\right) - \frac{3}{2}\Psi\left(\frac{(4/3)B_{so}+B_\phi}{B}+\frac{1}{2}\right) - \frac{1}{2}\ln\left(\frac{B_\phi}{B}\right) - \ln\left(\frac{B_{so}+B_e}{B}\right) + \frac{3}{2}\ln\left(\frac{(4/3)B_{so}+B_\phi}{B}\right)\right]. \quad (1)$$

Here, $\Psi(x)$ is the digamma function. Three subscripts are used to denote different scattering processes: $\phi$ for inelastic dephasing process, $so$ for spin-orbit scattering, and $e$ for elastic scattering. $B_{\phi,so,e}$ is the characteristic field for each corresponding scattering mechanism and is given by $B_{\phi,so,e} = \hbar/(4eL^2_{\phi,so,e})$. $L_\phi$ is the dephasing length, $L_{so}$ the spin relaxation length, and $L_e$ the mean free path. We fit our measured magnetoconductance data at different back gate voltages to Eq. (1). Figures 3b and 3c (black solid lines) show the results of fitting for the measured low-field magnetoconductance at $V_{bg} = 30$ V and $V_{bg} = -14$ V. Here, the fitting range is limited within $\pm 0.2$ T to satisfy the small field precondition. It is seen that the measured magnetoresistances can be well described by the HLN formula.

Figure 3d shows the extracted dephasing length $L_\phi$, spin relaxation length $L_{so}$ and mean free path $L_e$ as a function of back gate voltage. The main feature shown in Figure 3d is that $L_\phi$ is strongly dependent on back gate voltage $V_{bg}$, while $L_{so}$ shows a weak dependence on $V_{bg}$. The extracted $L_\phi$ decreases from 300 nm to 100 nm as $V_{bg}$ sweeps from 30 to -14 V, indicating that the dephasing process is stronger in the low carrier density region. In addition, $L_\phi$ is much larger than the thickness of the nanoplate at all back gate voltages, which is fully consistent with the 2D nature of transport in the $Bi_2O_2Se$ nanoplate we have assumed above. At the low temperature, the main source of dephasing is electron-electron interaction in two mechanisms. One is the Nyquist scattering representing the small energy transferred interactions of electrons with electromagnetic field fluctuations generated by the movement of neighboring electrons[34]. The other is the direct Coulomb interaction among the electrons. The observed enhanced dephasing process in our 2D $Bi_2O_2Se$ nanoplate system with decreasing electron density can be attributed to an enhancement in the Nyquist scattering. This is because the direct Coulomb interaction among the electrons is insensitive to the electron density in a 2D system. However, as the electron density decreases, the screening of cruising electrons reduces and thus the Nyquist scattering becomes stronger.

As seen in Figure 3d, the phase coherence length $L_\phi$ and the spin-orbit coupling length $L_{so}$ exhibit a crossover with decreasing electron density. Thus, the crossover between WAL and WL characteristics seen in Figure 3a can be attributed to the change in the relative values of $L_\phi$ and $L_{so}$. The extracted $L_{so}$ in the Bi$_2$O$_2$Se nanoplate is about 150 nm, which is shorter than that in Al$_x$Ga$_{1-x}$N/GaN 2DEG (~290 nm)[28] and InSb nanowires (~250 nm)[31]. Thus, as we expected, the Bi$_2$O$_2$Se nanoplate exhibits a strong spin-orbit interaction and is an excellent candidate for realizing a helical electron system and a topological superconducting device for topological quantum computation. Here, we would like to note that due to the inversion symmetry presented in the material system, the SOI observed in the nanoplate is most likely of the Rashba type and should be able to be tuned by applying an electric field across the nanoplate. The weak tunability of $L_{so}$ by the back gate voltage observed in Figure 3(d) arises from the fact that due to the ultra-thin Bi$_2$O$_2$Se conduction channel and a single back gate voltage in the device structure, the voltage applied to the back gate could only efficiently tune the carrier density in the nanoplate, but not an electric field across it. A way to achieve a largely tunable SOI Bi$_2$O$_2$Se device is to employ a dual-gate structure by, e.g., adding a top gate to the present device. Currently, such a dual-gate Bi$_2$O$_2$Se device technology is under development.

The WAL and WL characteristics of the magnetotransport in the Bi$_2$O$_2$Se nanoplate device are also studied at different temperatures. Figure 4a displays the measured low-field magnetoconductance of the device at $V_{bg} = 30\ V$ at temperatures ranging from 2 to 13 K. For clarity, the measured data at different temperatures are again vertically offset successively in this figure. At a low temperature, a sharp WAL characteristic magnetoconductance peak is seen in the vicinity of zero field. As temperature increases, the WAL peak is gradually suppressed and the low-field magnetoconductance evolves to show an overall WL dip, i.e., again a WAL-WL crossover behavior. The measured data at different temperatures are again fitted to the HLN formula of Eq. (1) and the results are presented by black solid lines in Figure 4a. Figure 4b shows the extracted characteristic transport lengths $L_\phi$ and $L_{so}$ from the fittings. It is shown that approximately $L_{so}$ does not change with increasing temperature. However, $L_\phi$ is strongly temperature dependent; it decreases with increasing temperature following a power law of $L_\phi \sim T^{-0.57}$, which is in a good agreement with our above assumption and analysis that the transport in the Bi$_2$O$_2$Se nanoplate is of the 2D nature and the dephasing dominantly arises from the electron-electron scattering processes with small energy transfers[34]. In correspondence to the temperature-driven WAL-WL crossover characteristics of the magnetoconductance, the relative values of $L_\phi$ and $L_{so}$ also show a crossover with increasing temperature. At low temperatures, the quantum correction to the low-field magnetotransport in $Bi_2O_2Se$ nanoplate is dominantly governed by the WAL characteristics originating from strong spin-orbit interaction (i.e., short spin-orbit interaction length). However, at high temperatures, the quantum correction to the low-field magnetotransport is dominantly governed by the WL characteristics due to thermally enhanced dephasing processes (i.e., short coherence length).

In conclusion, we have detected the presence of a strong spin-orbit interaction in an ultra-thin $Bi_2O_2Se$ nanoplate through the magnetotransport measurements at low temperatures. The low-field magnetoconductance of the nanoplate is measured at different carrier densities. The quantum correction to the magnetoconductance shows dominantly the WAL characteristics at high carrier densities and the WL characteristics at low carrier densities. The measurements are analyzed based on the HLN 2D diffusive transport theory and characteristic transport lengths, such as the phase coherence length $L_\phi$ and spin-relaxation length $L_{so}$, in the $Bi_2O_2Se$ nanoplate are extracted. It is shown that $L_{so}$ is relatively short (~150 nm) in the nanoplate, when compared with other semiconductor materials, representing a sign of strong spin-orbit interaction, and is roughly independent of the carrier density of the nanoplate. The magnetotransport measurements have also been performed for the $Bi_2O_2Se$ nanoplate at variable temperatures. The quantum correction to the low-field magnetoconductance again shows a WAL-WL crossover with increasing temperature. The extracted spin relaxation length is also rather independent of temperature. However, the extracted phase coherence length decreases with increasing temperature following the power law of $L_\phi \sim T^{-\alpha}$ with $\alpha \approx 0.5$, implying that the dephasing dominantly originates from the small energy-transferred electron-electron scattering processes. Our experimental study lays out a solid foundation for the further exploration of layered semiconducting $Bi_2O_2Se$ materials for possible applications in spintronics and topological quantum devices.

**Acknowledgements**

This work is supported by the Ministry of Science and Technology of China through the National Key Research and Development Program of China (Grant Nos. 2017YFA0303304, 2016YFA0300601, 2017YFA0204901, and 2016YFA0300802), the National Natural Science Foundation of China (Grant Nos. 91221202, 91421303, and 11274021). H.Q.X. also acknowledges financial support from the Swedish Research Council (VR).

**Author contributions**

H.Q. X. conceived and supervised the project. M.M. and S.H. fabricated the devices, carried out the electrical and magnetotransport measurements. C.T., J.W. and H.P. grew the materials. Y.J. participated in the device fabrication and measurements. M.M., S.H. and H.Q.X. analyzed the data and wrote the manuscript with contributions from all authors. All authors contributed to the discussion of the results and the interpretation of the experimental data acquired.

**Additional Information**

**Supplementary Information** accompanies this paper.

**Competing Interesting:** The authors declare no competing financial interests.

**Reprint and permission** information is available online.

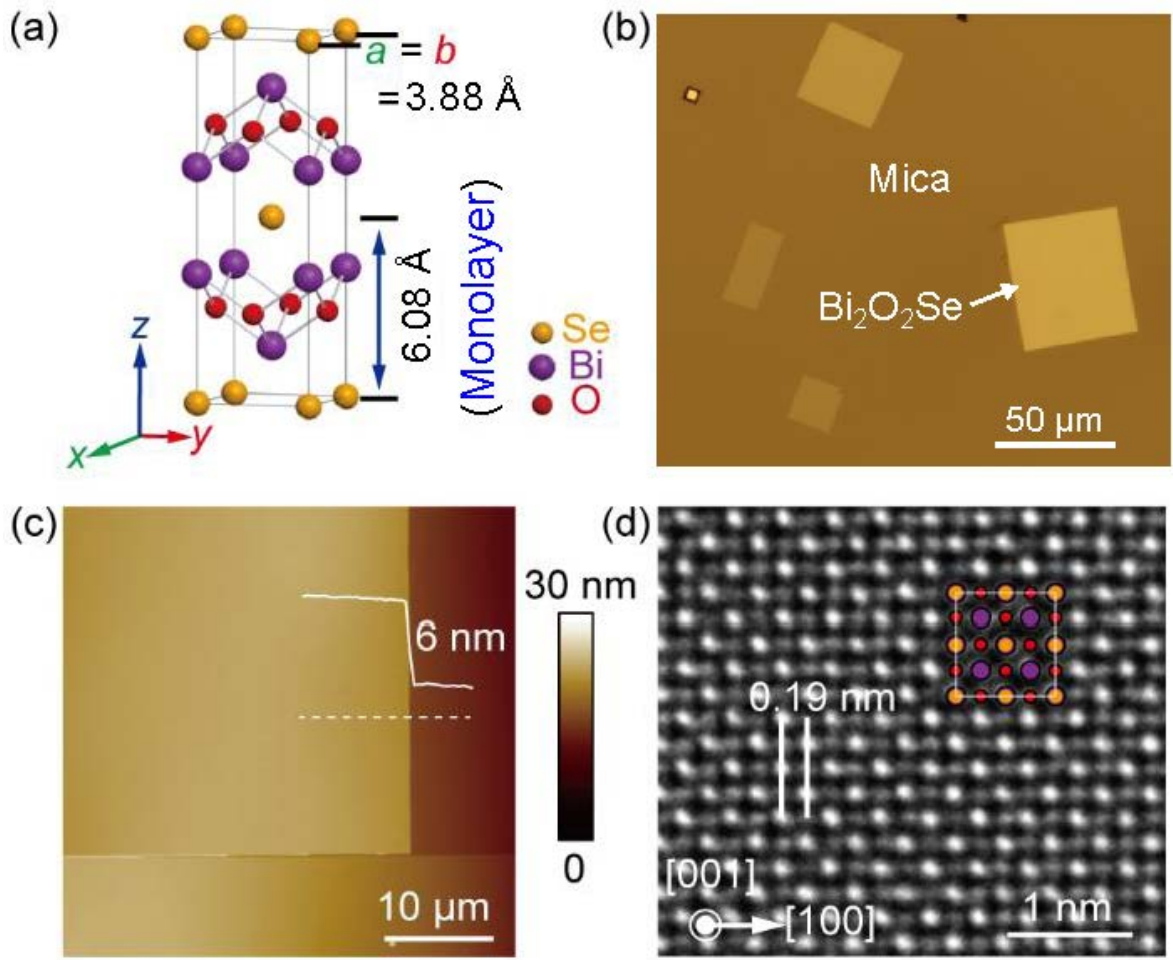

**Figure 1.** (a) Crystal structure of layered Bi$_2$O$_2$Se. (b) Representative optical image of a few as-grown Bi$_2$O$_2$Se nanoplates on a growth substrate mica. (c) AFM image of an as-grown Bi$_2$O$_2$Se nanoplate on the growth substrate. The nanoplate has an atomically flat surface and a thickness of 6 nm (see the white solid line for the height profile measured by AFM along the dashed line across an edge of the nanoplate). (d) HRTEM image of an as-grown Bi$_2$O$_2$Se nanoplate. A well-defined alternative arrangement of dark and bright lattices (with a spacing of ~0.19 nm) is observed. Marked in a square dark background region are atomic positions of atoms Bi (violet dots), O (red dots), and Se (yellow dots).

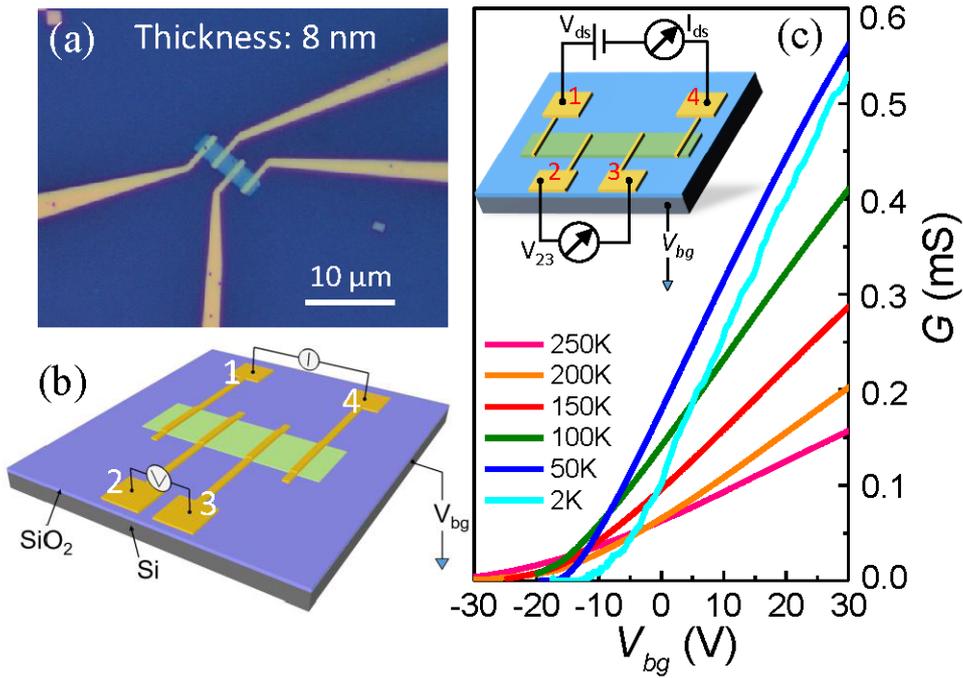

**Figure 2.** (a) Optical image of the fabricated $Bi_2O_2Se$ nanoplate device measured for this work. The device is made on a $SiO_2$/Si substrate. The nanoplate has a length of ~10 µm, a width of ~3 µm and a thickness of ~8 nm. The four finger Ti/Au contacts made on the nanoplate have a width of ~800 nm and a pitch of ~2.6 µm. The scale bar is 10 µm. (b) Titled schematic view of the device and circuit setup for the magnetotransport measurements using lock-in technique. Here, the contacts are labeled as 1 to 4 and the Si substrate is used as the back gage. (c) Conductance of the device measured in the setup as shown in the inset as a function of back gate voltage $V_{bg}$ at different temperatures. Here, the conductance $G=I_{ds}/V_{23}$ is plotted as a function of $V_{bg}$, where $I_{ds}$ is the current passing through the $Bi_2O_2Se$ nanoplate channel and $V_{23}$ is the voltage between contacts 2 and 3, measured with a constant source-drain bias voltage of $V_{ds} = V_{14} = 5$ mV applied between contacts 1 and 4.

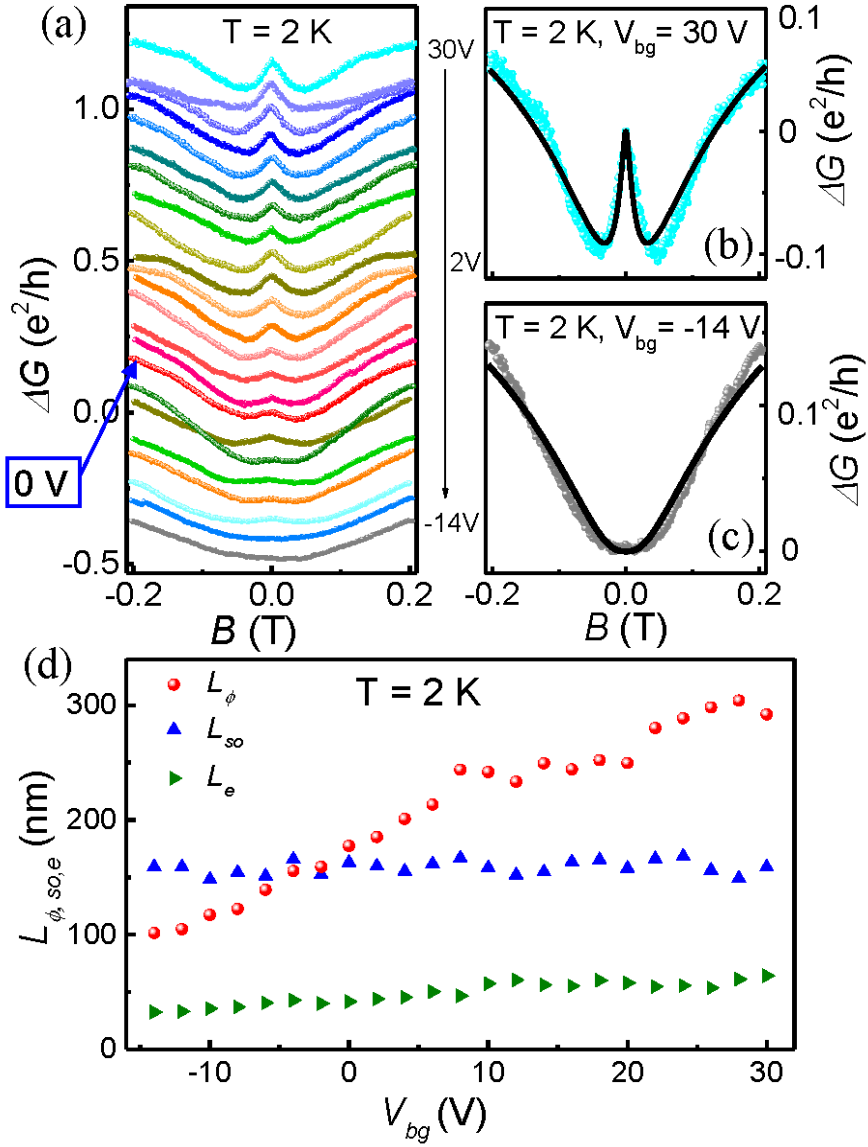

**Figure 3.** (a) Low-field magnetoconductance measured for the $Bi_2O_2Se$ nanoplate device at various back gate voltage and a temperature of 2 K. Here, the curve measured at $V_{bg} = 0$ V is marked by a blue arrow and all other curves are successively vertically offset for clarity. The measurements show that a crossover from WAL to WL occurs when $V_{bg}$ is swept from 30 to -14 V. (b) and (c) The same low-field magnetoconductance measurement data as in (a) for $V_{bg} = 30$ V and $V_{bg} = -14$ V. The black solid lines are theoretical fits to the experimental data. (d) Dephasing length $L_\phi$, spin relaxation length $L_{so}$, and mean free path $L_e$ extracted for the device as a function of back gate voltage $V_{bg}$ from the measured data shown in (a).

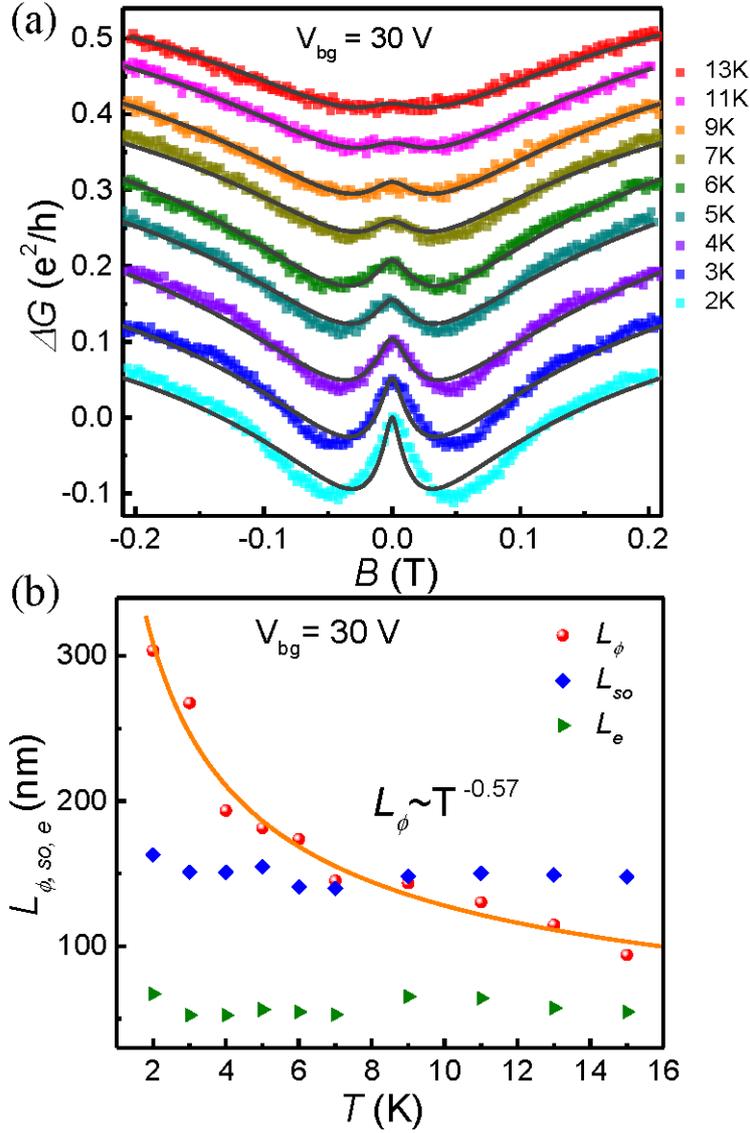

**Figure 4.** (a) Low-field magnetoconductance of the $Bi_2O_2Se$ nanoplate device measured at a fixed back gate voltage of $V_{bg}$ = 30 V but different temperatures. The lowest curve is for the measurements at 2 K and all other curves are successively vertically offset for clarity. The black solid lines are theoretical fits to the experimental data. Here, a WAL–WL crossover is seen to occur with increasing temperature. (b) Dephasing length $L_\phi$, spin relaxation length $L_{so}$, and mean free path $L_e$ extracted for the device as a function of temperature $T$ from the measured data shown in (a). The orange solid line is the power-law fit to the extracted dephasing length $L_\phi$, showing $L_\phi \sim T^{-0.57}$.

# Supplementary Materials for
# Strong spin-orbit interaction and magnetotransport in semiconductor $Bi_2O_2Se$ nanoplates

Mengmeng Meng,[1] Shaoyun Huang,[1,*] Congwei Tan,[2] Jinxiong Wu,[2] Yumei Jing,[1] Hailin Peng[2,*] and H. Q. Xu[1,3,*]

[1]*Beijing Key Laboratory of Quantum Devices, Key Laboratory for the Physics and Chemistry of Nanodevices, and Department of Electronics, Peking University, Beijing 100871, China*

[2]*Center for Nanochemistry, Beijing National Laboratory for Molecular Sciences (BNLMS), College of Chemistry and Molecular Engineering, Peking University, Beijing 100871, China.*

[3]*Division of Solid State Physics, Lund University, Box 118, S-221 00 Lund, Sweden*

[*]Corresponding authors: Professor H. Q. Xu (hqxu@pku.edu.cn), Dr. Shaoyun Huang (syhuang@pku.edu.cn), and Professor Hailin Peng (hlpeng@pku.edu.cn)

## Contents

## S1. Extraction of the carrier density $n_{2D}$ and the field-effect mobility $\mu$

To extract the carrier density in a Bi$_2$O$_2$Se nanoplate, we employ the parallel plate capacitor model, which is a good approximation when the size of the nanoplate is much larger than the thickness of the dielectric. Based on this parallel plate capacitor model, the capacitance ($C$) can be evaluated from

$$C = \frac{\varepsilon_0 \varepsilon_r A}{d}, \qquad (1)$$

where $\varepsilon_0$ is the electric permittivity of vacuum, $\varepsilon_r$ is the relative permittivity of SiO$_2$ ($\varepsilon_r = 4$), $A$ is the area of the nanoplate, and $d$ is the thickness of the dielectric SiO$_2$. The electric charge ($Q$) presented in the nanoplate is given by

$$Q = C \cdot \Delta V = \frac{\varepsilon_0 \varepsilon_r A}{d} \cdot \Delta V, \qquad (2)$$

where

$$\Delta V = V_{bg} - V_{th}, \qquad (3)$$

with $V_{bg}$ being a voltage applied to the back gate and $V_{th}$ the threshold pinch-off back gate voltage at which no conduction carriers are present in the nanoplate at zero temperature (or after neglecting thermal excited carriers at a finite temperature). The carrier density $n_{2D}$ in the nanoplate is then obtained from

$$n_{2D} = \frac{Q}{eA} = \frac{\varepsilon_0 \varepsilon_r}{ed} \cdot (V_{bg} - V_{th}) = \frac{C_{gs}}{e} \cdot (V_{bg} - V_{th}), \qquad (4)$$

where $e$ is the elementary charge and $C_{gs} = C/A$ is the capacitance per unit area.

The field-effect mobility is extracted from the linear region of the transfer characteristic curve of the Bi$_2$O$_2$Se nanoplate device by the following equation,

$$\mu = \frac{1}{C_{gs}} \cdot \frac{L}{W} \cdot \frac{1}{V_{ds}} \cdot \frac{dI_{ds}}{dV_{bg}}, \qquad (5)$$

where $L$ and $W$ are the channel length (i.e., the distance between the source and drain contacts) and the channel width (i.e., the width of the nanoplate), $V_{ds}$ is a small source-drain voltage applied, and $\frac{dI_{ds}}{dV_{bg}}$ is the transconductance taken from the linear region of the transfer characteristics of the device.

## S2. Extraction of the mean free path $L_e$ from carrier density $n_{2D}$ and mobility $\mu$

In a conduction channel, the mean free path $L_e$ describes the mean distance that electrons travel between two collisions with scattering centers and phonons, and is obtained from

$$L_e = v_F \tau \ , \tag{6}$$

where $\tau$ is the momentum relaxation time, i.e., the time electrons travel between two collisions with scattering centers and phonons, and $v_F$ is the Fermi velocity of the carriers. In the effective mass approximation, the Fermi velocity is given by

$$v_F = \frac{\hbar k_F}{m^*} = \frac{\hbar}{m^*}\sqrt{2\pi n_{2D}} \ , \tag{7}$$

where $\hbar$ is the reduced Planck constant, $k_F$ is the Fermi wave vector, and $m^*$ is the effective mass of the carriers. The momentum relaxation time $\tau$ can be evaluated from the mobility through the relation

$$\mu = \frac{e\tau}{m^*} \ . \tag{8}$$

Thus, the mean free path can be obtained from

$$L_e = \frac{\hbar \mu}{e}\sqrt{2\pi n_{2D}} \ . \tag{9}$$

## S3. Transfer characteristics of the Bi$_2$O$_2$Se nanoplate field-effect device at 2 K

In this supplementary note, we present in Figure S1 the transfer characteristics of the Bi$_2$O$_2$Se nanoplate field-effect device measured at 2K and describe, as an example, the procedure in which the carrier density, the field effect mobility, and the mean free path in the device are extracted.

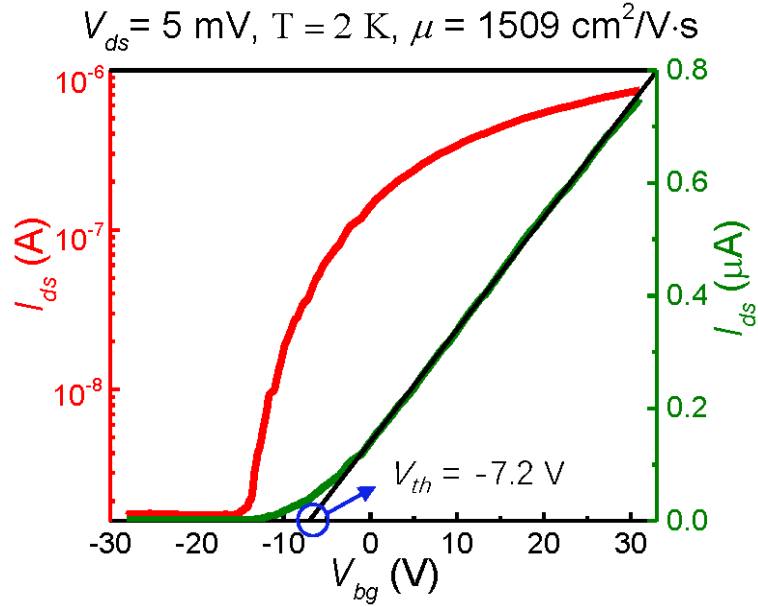

**Figure S1.** Source-drain current $I_{ds}$ vs. back-gate voltage $V_{bg}$ (transfer characteristics) measured for the Bi$_2$O$_2$Se nanoplate device discussed in the main article at a source-drain bias voltage of $V_{ds}$= 5 mV and a temperature of $T$=2 K, see the inset of Figure 2c in the main article for the measurement setup. The green curve is the plot for $I_{ds}$ in linear scale, while the red curve is the plot for $I_{ds}$ in logarithmic scale. The channel is of n-type. The black solid line is the linear fit to the transfer characteristics. The fitting line intersects with the horizontal axis, giving a channel pinch-off threshold voltage of $V_{th}$ = -7.2 V. With this threshold voltage and the estimated capacitance per unit area $C_{gs} = 1.18 \times 10^{-4}$ F/m$^2$, the carrier density in the nanoplate can be determined for $V_{bg}$ in the region where the transfer characteristics show a good linear dependence on $V_{bg}$. The slope of the fitting line, together with the channel length $L$=7.8 µm, the channel width $W$=3 µm, and $C_{gs}$= $1.18 \times 10^{-4}$ F/m$^2$, gives the field-effect mobility of $\mu$= 1509 cm$^2$/V·s in the Bi$_2$O$_2$Se nanoplate channel at 2 K. Using this mobility value and the carrier density extracted as a function of $V_{bg}$, the mean free path in the Bi$_2$O$_2$Se nanoplate channel at 2K can be estimated. The same procedure has been employed to extract the mobility, the carrier density and the mean free path for the device from the transfer characteristics measured at other temperatures.

## S4. Carrier density $n_{2D}$, mobility $\mu$, and mean free path $L_e$ extracted for the $Bi_2O_2Se$ nanoplate discussed in the main article at different temperatures

In this supplementary note, the carrier density $n_{2D}$, mobility $\mu$, and mean-free path $L_e$ extracted for the $Bi_2O_2Se$ nanoplate studied in the main article as a function of back-gate voltage $V_{bg}$ at different temperatures, using the same procedure as described in the caption to Figure S1, are presented.

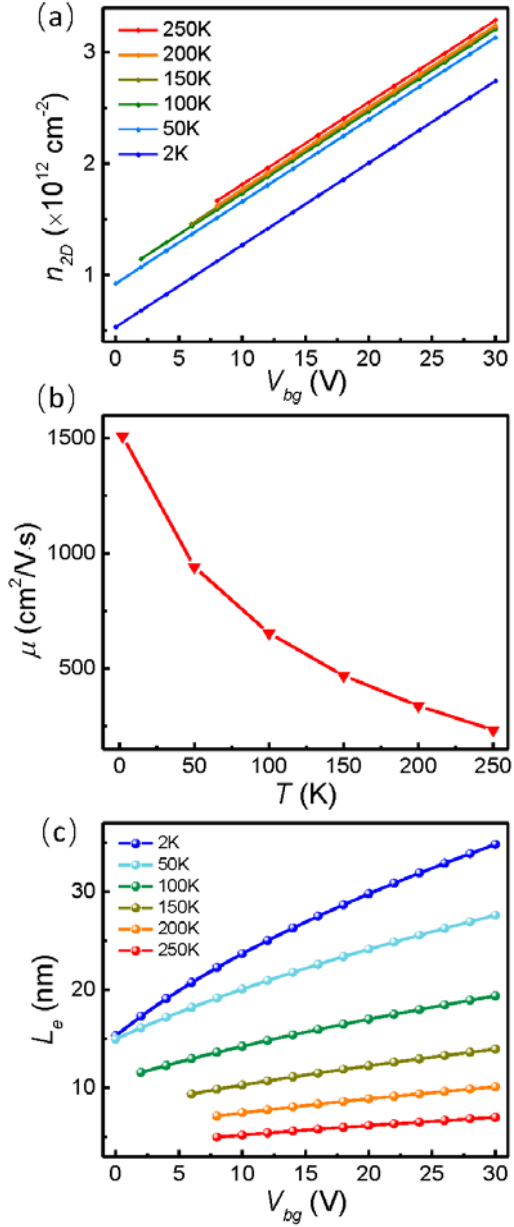

**Figure S2.** (a) Sheet carrier density $n_{2D}$ as a function of back gate voltage $V_{bg}$ at different temperatures, (b) carrier mobility $\mu$ as a function of temperature, and (c) carrier mean free path $L_e$ as a function of $V_{bg}$ at different temperatures in the same $Bi_2O_2Se$ nanoplate as studied in the main article, extracted using the same procedure as described in the caption to Figure S1.